\documentclass[12pt]{spieman}  

\usepackage{amsmath,amssymb}
\usepackage{xspace}
\usepackage{CJK}
\usepackage{graphicx}
\usepackage[numbers,sort&compress]{natbib} 

\usepackage{hyperref}
\hypersetup{
    colorlinks=true,
    linkcolor=blue,
    filecolor=magenta,      
    urlcolor=cyan,
}

\usepackage{color}

\newcommand{\Lsol}{\ensuremath{\rm{L}_\odot}\xspace}
\newcommand{\um}{\ensuremath{\mu\rm{m}}\xspace}

\newcommand{\kms}{\ensuremath{\rm{km\,s}^{-1}}\xspace}
\newcommand{\wmt}{\ensuremath{\rm{W\,m}^{-2}}\xspace}

\newcommand{\lir}{\ensuremath{L_{\rm{IR}}}\xspace}

\newcommand{\cii}{[C{\scriptsize II}]\xspace}
\newcommand{\oi}{[O{\scriptsize I}]\xspace}
\newcommand{\nii}{[N{\scriptsize II}]\xspace}
\newcommand{\niii}{[N{\scriptsize III}]\xspace}

\newcommand{\oiii}{[O{\scriptsize III}]\xspace}
\newcommand{\siii}{[S{\scriptsize III}]\xspace}
\newcommand{\neii}{[Ne{\scriptsize II}]\xspace}
\newcommand{\neiii}{[Ne{\scriptsize III}]\xspace}

\newcommand{\etal}{et~al.\xspace}

\newcommand{\arc}{\ensuremath{''}\xspace}

\newcommand{\vout}{\ensuremath{v_{\mathrm{out}}}\xspace}

\newcommand{\aap}{Astron. \& Astrophys.}
\newcommand{\apj}{Astrophys. J.}
\newcommand{\apjl}{Astrophys. J. Letters}
\newcommand{\apjs}{Astrophys. J. Supp.}
\newcommand{\pasa}{Publ. Astron. Soc. Aus.}
\newcommand{\araa}{Ann. Rev. Astron. Astrophys.}
\newcommand{\mnras}{Mon. Not. R. Astron. Soc.}
\newcommand{\nat}{Nature}
\newcommand{\aapr}{Astron. \& Astrophys. Rev.}

\title{High-Redshift Extragalactic Science with the Single Aperture Large Telescope for Universe Studies (\textit{SALTUS}) Space Observatory}

\author[a,*]{Justin Spilker}
\author[b,\dag]{Rebecca C. Levy}
\author[b]{Daniel Marrone}
\author[b]{Stacey Alberts}
\author[c,d,e]{Scott C. Chapman}
\author[f]{Mark Dickinson}
\author[b]{Eiichi Egami}
\author[g]{Ryan Endsley}
\author[h]{Desika Narayanan}
\author[b]{George Rieke}
\author[i]{Antony A. Stark}
\author[j,k]{Alexander Tielens}
\author[b]{Christopher K. Walker}
\affil[a]{Department of Physics and Astronomy and George P. and Cynthia Woods Mitchell Institute for Fundamental Physics and Astronomy, Texas A\&M University, 4242 TAMU, College Station, TX 77843-4242}
\affil[b]{University of Arizona, Department of Astronomy and Steward Observatory, Tucson, AZ 85719}
\affil[c]{Department of Physics and Astronomy, University of British Columbia, Vancouver, British Columbia, Canada}
\affil[d]{National Research Council, Herzberg Astronomy and Astrophysics, Victoria, British Columbia, Canada}
\affil[e]{Department of Physics and Atmospheric Science, Dalhousie University, Halifax, Nova Scotia, Canada}
\affil[f]{NSF’s NOIRLab, Tucson, AZ 85719, USA}
\affil[g]{Department of Astronomy, University of Texas at Austin, Austin, TX 78712}
\affil[h]{Department of Astronomy, University of Florida,
Gainesville, FL, USA}
\affil[i]{Harvard-Smithsonian Center for Astrophysics, 60 Garden Street, Cambridge, MA 02138, USA}
\affil[j]{Leiden Observatory, P.O. Box 9513, Leiden, The Netherlands}
\affil[k]{Astronomy Department, University of Maryland, College Park, MD
20742, USA}

\begin{document} 
\maketitle

\begin{abstract}
This paper presents an overview of the high-redshift extragalactic science case for the \textit{Single Aperture Large Telescope for Universe Studies (SALTUS)} far-infrared NASA probe-class mission concept. Enabled by its 14m primary reflector, SALTUS offers enormous gains in spatial resolution and spectral sensitivity over previous far-IR missions. SALTUS would be a versatile observatory capable of responding to the scientific needs of the extragalactic community in the 2030s, and a natural follow-on to the near- and mid-IR capabilities of JWST. Key early-universe science goals for SALTUS focus on understanding the role of galactic feedback processes in regulating galaxy growth across cosmic time, and charting the rise of metals and dust from the early universe to the present.  This paper summarizes these science cases and the performance metrics most relevant for high-redshift observations.
\end{abstract}


{\noindent \footnotesize\textbf{*}Justin Spilker,  \linkable{jspilker@tamu.edu}}

{\noindent \footnotesize\textbf{\dag}National Science Foundation Astronomy \& Astrophysics Postdoctoral Fellow}

\begin{spacing}{1.1}   

\section{Introduction} \label{intro} 

The Single Aperture Large Telescope for Universe Studies (SALTUS) observatory is a far-infrared space mission concept proposed to NASA under the recent APEX call for proposals (Chin \etal, subm.). SALTUS would provide orders-of-magnitude gains in sensitivity throughout the far-infrared wavelength range ($\approx$30--700\,\um), enabled by sensitive instrumentation and a 14\,m primary mirror. The large aperture size would also provide $\sim$1\arc spatial resolution at these wavelengths, solving the spatial confusion problems that have plagued past far-IR space observatories. The two instruments planned for SALTUS are HiRX, a multi-pixel, multi-band heterodyne receiver system (Silva \etal, subm.), and SAFARI-Lite, a direct-detection grating spectrometer providing simultaneous 35--230\,\um spectroscopy (Roelfsema \etal, subm.).  The overall telescope and spacecraft architectures are described in a series of papers elsewhere in this issue (Arenberg \etal, Kim \etal, Harding \etal subm.).

This paper provides an overview of the promise of SALTUS for high-redshift extragalactic science. Accompanying papers describe the plans for guaranteed-time and guest observing (Chin \etal), SALTUS' contributions to Milky Way and nearby galaxies science (Levy \etal), star and planet formation (Schwarz \etal), and solar system observations (Anderson 
\etal). In Section~\ref{scidiag} we give a high-level overview of the key observable features that enable SALTUS' high-redshift science goals. Section~\ref{perf} briefly summarizes the most relevant SALTUS performance characteristics for high-redshift observing programs. 
Section~\ref{sci} highlights several key high-redshift measurements that would address key open questions in the field. We conclude in Section~\ref{conclusions}.

\section{Connecting Far-IR Observables to Science Outcomes} \label{scidiag}

The mid- and far-infrared wavelength ranges are unique in the extraordinarily wide range of astrophysical phenomena accessible by atomic and molecular spectral diagnostics. Spectral features trace gas over many orders of magnitude in temperature and density, from cold molecular gas probed by far-IR molecular lines to hot gas traced by high-ionization lines typically associated with black hole accretion. In this section we provide a (very) brief overview of the various diagnostics key to SALTUS' high-redshift science case. More detailed descriptions can be found in many previous works, including the study reports for the SPICA and Origins Space Telescope mission concepts \citep{roelfsema18,meixner19}.

A key advantage to all of these diagnostics is the near total insensitivity to dust extinction. With the exception of silicate absorption features at rest-frame 9.7 and 18\,\um, dust extinction is negligible throughout the mid- and far-IR, more than an order of magnitude lower than at optical wavelengths and  \citep[e.g.][]{chiar06,wang19}. The gas column density $N_H$ likewise does not affect far-IR measurements, in contrast to the strong absorption of X-rays due to hydrogen along the line of sight.

The mid-/far-IR wavelength range uniquely allows simultaneous measurements of both the star formation rate (SFR) and black hole accretion rate (BHAR) of galaxies through a suite of ionic fine-structure lines. Ions with relatively low ionization potential (e.g. \neii, \neiii, \siii) are robust tracers of HII regions surrounding young massive stars. More highly-ionized species (e.g. [O{\scriptsize IV}], [Ne{\scriptsize V}]) are predominantly produced by the hard UV radiation fields surrounding AGN, providing direct BHAR tracers. Cooler ionized and atomic gas in the transition region between massive stars and their parent molecular clouds are traced by the brightest ISM cooling lines, including \cii, \oi, \oiii, and \nii. Together, these lines provide a complete picture of the interstellar medium of galaxies.

The ratios of these bright spectral lines also allow robust estimates of the metallicity and abundance patterns in galaxies \citep[e.g.][]{pereirasantaella17}. Line ratios of \niii/\oiii, for example, trace the [N/O] abundance pattern \citep{peng21}, which can either be combined with direct tracers of the H abundance (e.g. free-free emission) or used to estimate galaxy metallicities alone via the correlation between [N/O] and [O/H] \citep[e.g.][]{spinoglio22,chen23}. Though studied in far fewer galaxies than the typical strong lines in the optical, these IR diagnostics will prove especially useful in the dusty galaxies that dominate the total SFR density over the last 12\,Gyr of cosmic evolution \citep{chartab22}.

Signatures of galactic feedback processes are also present in IR wavelengths. In spatially-unresolved data, massive galactic outflows are typically recognized by excess flux in the high-velocity wings of spectral lines. While X-ray diagnostics probe only the very hottest and low-density material in outflows, and optical diagnostics provide access to warm ionized gas outflows, the IR is unique in its ability to probe outflowing gas across many orders of magnitude in temperature and density. From the coldest molecular phases that dominate the total outflowing mass (lines of OH, H$_2$O; \cite{gonzalezalfonso14,spilker20}), to cool neutral gas (\cii line wings; \cite{herreracamus21}), to warm ionized gas (\oiii), a comprehensive census of the gas in galactic winds can uniquely be assembled solely using IR spectroscopy.

At the intersection between few-atom molecules and macroscopic dust grains in the ISM lie Polycyclic Aromatic Hydrocarbons (PAHs), complex organic molecules consisting of hundreds or thousands of atoms. Often thought to represent the small-grain tail of the dust grain size distribution, PAHs have very bright rest-frame mid-IR bands that represent up to 20\% of the total IR luminosity \citep{smercina18} and are thus detectable to large distances. Furthermore, the ratios between various PAH features allow estimates of the grain size distribution, PAH ionization fraction, and ambient radiation field strength \citep{draine21}.

With PAHs tracing, in effect, the small-size end of the dust grain size distribution, larger grains emit long-wavelength continuum in accordance with their temperature. Hundreds of thousands of dusty galaxies are known (most without spectroscopic redshift confirmation) thanks to wide-area surveys with Herschel and ground-based telescopes like the South Pole Telescope. While these imaging surveys often face challenges due to confusion, SALTUS' larger aperture allows the continuum of individual galaxies to be recovered from its wide-band spectrometer (Sec.~\ref{conf}). Dust grains can also be probed using the well-known silicate features at 9.7 and 18\,\um, which appear prominently in absorption against the bright continuum of AGN. 

Finally, the far-IR also provides access to direct and indirect tracers of molecular hydrogen, H$_2$. The rotational H$_2$ lines, particularly those at 17.0 and 28.2\,\um, are tracers of warm molecular gas ($\sim$500-1000\,K) in conditions often reached in shocks. Because H$_2$ itself lacks a permanent dipole moment, it cannot trace colder molecular gas. Its deuterated counterpart HD, however, can, providing robust estimates of the total molecular gas contents of galaxies because the cosmic deuterium abundance is well known.

\section{SALTUS Performance for High-Redshift Science} \label{perf}

The design principles of the spacecraft, telescope, and instrumentation for SALTUS are described in a series of papers in this issue. Far more detailed information can be found in those works, but here we provide a brief summary of the SALTUS performance metrics as they apply to common needs of the high-redshift observational community. While some spectral lines will be accessible with the heterodyne HiRX receiver at significant lookback times, especially in very IR-luminous sources, most high-redshift science is likely to be carried out using the SAFARI-Lite grating spectrometer instead. We focus here on the SALTUS/SAFARI-Lite performance as relevant for high-redshift science (Roelfsema \etal, subm.). 

Throughout this section, we compare to both past and present IR missions (JWST, Herschel), as well as to the now-cancelled SPICA mission concept. SPICA was intended to be an ESA-led mission consisting of a 2.5\,m primary mirror cryogenically cooled to $<$8\,K. We include this comparison to illustrate the contrasts between these two very different mission architectures: SPICA, with a small but very cold primary mirror, and SALTUS, with a much larger primary reflector passively cooled to $<$40\,K.

\subsection{Wavelength Coverage}

The SAFARI-Lite instrument will provide simultaneous coverage from 34--230\,\um at a spectral resolution $R\approx300$ ($\Delta v \approx 1000$\,\kms). Four co-aligned bands from each of six spatial pixels are dispersed onto 180-detector MKID arrays. This wavelength regime covers the entire rest-frame far-infrared at $z=0$. During the continued operations of JWST/MIRI and supplemented at longer wavelengths by ALMA, these instruments combined will provide nearly uninterrupted access to the entire infrared/submillimeter/millimeter wavelength range. 

At higher redshifts, varying bright far-IR fine structure lines redshift into and out of the SAFARI-Lite bandpass (Figures~\ref{fig:specdpm}, \ref{fig:specjs}). Notably, \cii 158\,\um is accessible to $z<0.45$, \oi 63\, \um to $z<2.6$, \oi 145\, \um to $z<0.6$, \oiii 88\,\um to $z<1.6$, \oiii 52\,\um to $z<3.4$, and [Si{\scriptsize II}] 34.8\,\um to $z<5.6$. Conversely, other bright rest-frame mid-IR features redshift into the bandpass, including \siii 33.5\,\um for $z>0.01$ and bright lines of \neii 12.8\,\um and \neiii 15.5\,\um for $z \gtrsim 1.3$. High-ionization tracers of the BHAR, including [Ne{\scriptsize V}] 14.3, 24.3\,\um and [O{\scriptsize IV}] 25.9\,\um are similarly accessible for $z>0.3$. Many galaxies also show very bright and broad mid-IR bands generally attributed to Polycyclic Aromatic Hydrocarbons (PAHs), complex organic molecules containing dozens or hundreds of carbon atoms that are in some ways intermediate between simple molecules and macroscopic dust grains. The brightest of the PAH features, at 6.2, 7.7, and 11.2\,\um, redshift into the SAFARI-Lite bandpass for $z \gtrsim 2$. Finally, the low-lying H$_2$ rotational transitions at 28.2, 17.0\,\um redshift into the SAFARI-Lite bandpass beginning at $z=0.2$.

For high-redshift science, the broad and simultaneous wavelength coverage provided by SAFARI-Lite would allow the same spectral diagnostics to be used over a very wide range in wavelength, reducing systematic uncertainties inherent to comparing disparate diagnostics. By filling the gap between JWST/MIRI and ALMA, SALTUS would enable comprehensive and uniform analysis of large galaxy samples across cosmic time using the same observables.

\subsection{Point-Source Sensitivity}

Thanks to its large 14\,m aperture, SAFARI-Lite on SALTUS would offer transformational gains in sensitivity over past far-IR missions. The most directly comparable past instrument was the Photodetector Array Camera and Spectrometer (PACS) on the Herschel Space Observatory. PACS provided $R \sim 1000-3000$ integral field spectroscopy over a relatively narrow far-IR bandwidth ($\approx$800-3000\,\kms), and could reach typical 5$\sigma$, 1-hour line sensitivity $\approx 5 \times 10^{-18}$\,\wmt. In contrast, SALTUS/SAFARI-Lite reaches 5$\sigma$, 1-hour line sensitivity limits $<5 \times 10^{-20}$\,\wmt over the full 34--230\,\um wavelength range simultaneously, improving to $3-5 \times 10^{-21}$\,\wmt at the short-wavelength end of the bandpass. 

\begin{figure}
\begin{center}
\includegraphics[width=0.49\textwidth]{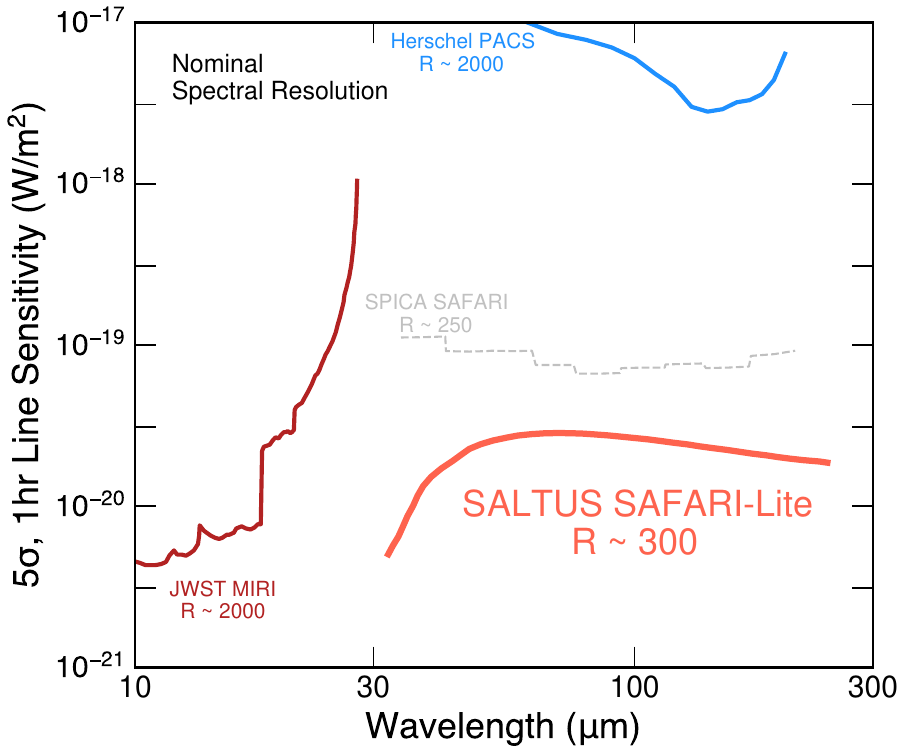}
\includegraphics[width=0.49\textwidth]{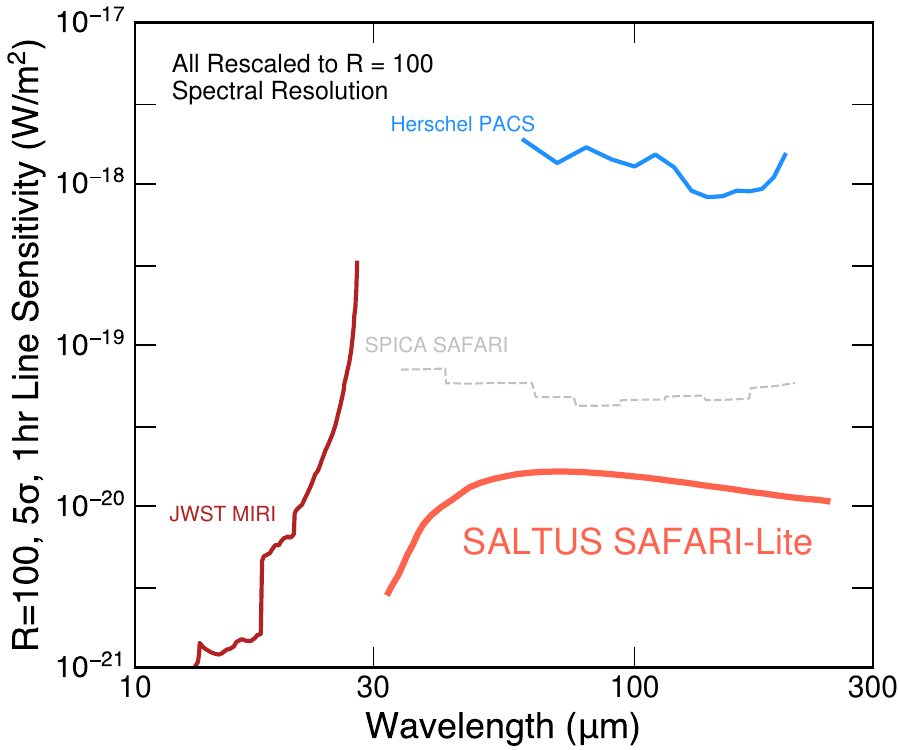}
\end{center}
\caption{ 
Comparison of the point-source line sensitivities for SALTUS/SAFARI-Lite with other past and existing missions. The left plot shows the instrument sensitivities at their nominal spectral resolution, while the right version rescales all sensitivity curves to a common $R=100$ spectral resolution to facilitate direct comparisons. This spectral resolution is appropriate for broad features such as redshifted PAH bands. Thanks to its large 14\,m aperture, SALTUS would provide extremely sensitive spectroscopy throughout the far-IR, reaching 2\,dex deeper than Herschel and $\approx$0.5\,dex deeper than even the cryogenic 2.5\,m SPICA mission concept.
}\label{fig:senscomparison}
\end{figure}

In Figure~\ref{fig:senscomparison} we compare the point-source sensitivity of SALTUS/SAFARI-Lite with other similar instruments, including JWST/MIRI MRS and Herschel/PACS. We also include the far-IR grating spectrometer instrument planned for the now-cancelled SPICA mission concept SAFARI.\footnote{SALTUS/SAFARI-Lite is a scaled-back version of SPICA/SAFARI that removes certain observing modes.}
This comparison is somewhat complicated by the varying spectral resolution of each instrument. For broad spectral features such as PAH emission, the sensitivity improves to lower spectral resolution as $R^{1/2}$ (because additional line flux is forced into fewer spectral channels), while for narrow lines it worsens (because wider channels average in additional zero-flux wavelengths and are poorly-matched to intrinsically narrow features).

Fig.~\ref{fig:senscomparison} makes clear that SALTUS/SAFARI-Lite would offer an enormous $>$100$\times$ gain in spectral sensitivity over Herschel/PACS. Interestingly, the instrument would also provide substantially better point-source sensitivity than SPICA  despite the fact that SPICA was planned to have a 2.5\,m primary mirror cryogenically cooled to $<$8\,K, in comparison to SALTUS' passively-cooled $<$40\,K reflector. The difference arises from the much larger collecting area of SALTUS compared to this smaller, colder mission. In other words, the 30$\times$ larger SALTUS collecting area more than compensates for the additional thermal background emission from the warmer aperture.

\subsection{Angular Resolution and (Lack of) Confusion Limits}\label{conf}

\begin{figure}
\begin{center}
\includegraphics[width=0.75\textwidth]{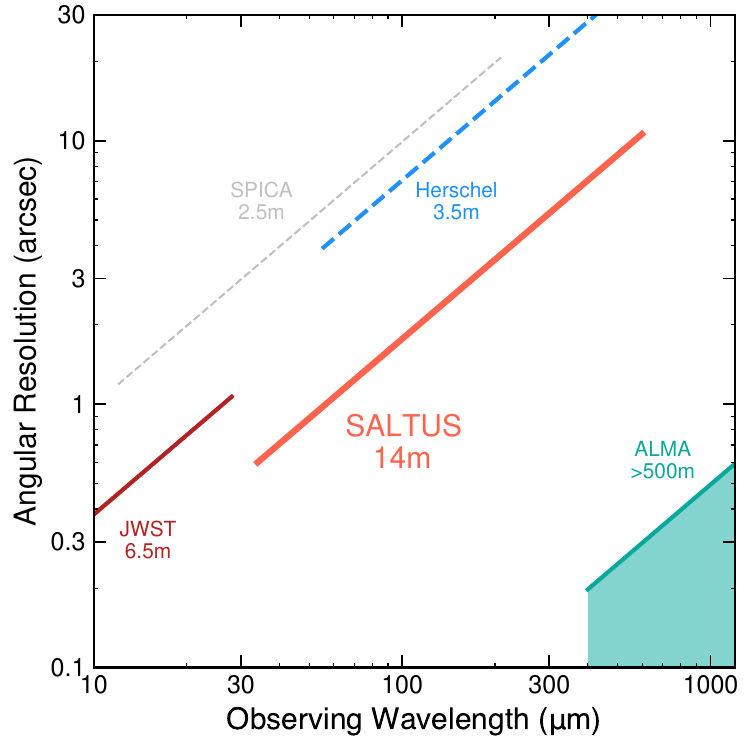}
\end{center}
\caption{ 
SALTUS is expected to reach diffraction-limited performance over its entire wavelength range, enabling $\approx$1\arc angular resolution in the far-IR. This would be a substantial improvement over existing or other possible facilities, approximately matching the resolution achievable with JWST at shorter wavelengths and ALMA at longer wavelengths.
}\label{fig:resolution}
\end{figure}

SALTUS is expected to reach diffraction-limited angular resolution over its entire wavelength coverage. As a result, SALTUS will offer large gains in resolution compared to Herschel ($16\times$ smaller beam area) or smaller $\sim$2\,m mission concepts (more than $50\times$ smaller beam area), reaching $\approx$1\arc resolution throughout the far-IR. This will allow nearby galaxies to be spectroscopically mapped in fine detail, roughly matching the resolution of JWST at mid-IR wavelengths and compact ALMA configurations at long wavelengths. The resolution of SALTUS is illustrated in Fig.~\ref{fig:resolution}.

In general, even 1\arc resolution will not significantly resolve high-redshift extragalactic targets, which are typically far more compact than local galaxies. The key gain for high-redshift science that SALTUS' high resolution provides is the complete lack of source confusion (see also Chin \etal, subm.). Confusion, caused by the collective emission from many unresolved galaxies within a single telescope beam, has long plagued far-IR observatories including Herschel. Confusion sets a fundamental limit to the faintest object that can be detected in a blind survey \citep[e.g.][]{dole04b}. Reaching beyond this limit requires knowing properties of faint sources \textit{a priori} (e.g. positions, redshifts, etc.) in order to deblend the confused emission.

Confusion was especially problematic for Herschel at long wavelengths, because the telescope spatial resolution was poor (36\arc at 500\,\um) and only 2D imaging data were available. This limit led to the development of deblending techniques that cross-matched with known galaxies from multiwavelength catalogs (e.g. XID/XID$+$; \cite{roseboom10,hurley17,liu18}). 

Confusion noise also limits even far-IR spectrometers for small aperture sizes, even with redshift information as a third dimension to help deblend the confused signal. The flagship-class ($\approx$\$ 10B) Origins Space Telescope mission concept, for example, found that spectral confusion limited the planned science goals for apertures $<$3.0\,m \citep[][Appendix E.1]{meixner19}. Even with its planned 5.9\,m aperture, only 40--70\% of sources were recovered by different attempts at extraction from a known input catalog. Worse still, line fluxes -- the key observable driving the mission -- were poorly recovered, with flux errors approaching a full order of magnitude in some cases. 

The problem of confusion is likely to be just as challenging for smaller apertures, but a detailed assessment as a function of aperture size is required. On one hand, a small aperture has a much larger beam area (scaling as D$^{-2}$), so even more objects will fall within a single telescope beam. On the other, a smaller aperture is less sensitive (instrumental noise also scaling with the collecting area D$^{-2}$), so perhaps faint confused emission would not have driven the mission design and goals in the first place. The confusion limit and impact on science outcomes is determined by the shape of the galaxy number counts and line luminosity functions as a function of redshift.

Smaller apertures require that the position and redshift of every galaxy in the field be known in advance, which has two main consequences: (1) the continuum emission of galaxies within the beam cannot be recovered reliably without severe assumptions, and (2) genuine discovery space is closed off, because deblending can only be performed for previously-identified galaxies.

The $<$3\arc spatial resolution provided by SALTUS completely eliminates source confusion thanks to the $\approx$16$\times$ smaller beam area compared to Herschel, or $\approx$50$\times$ smaller beam area compared to a 2\,m-class telescope. Aside from providing surveys with spatial resolution roughly matched to those currently being conducted with JWST/MIRI, SALTUS will provide the first confusion-free extragalactic surveys in the far-IR.

\subsection{Spectral Survey Speed}

The large primary mirror of SALTUS leads to both high sensitivity and high spatial resolution, well-matched to the modest-area surveys currently being conducted with JWST. Slewing and settling the large aperture, however, means that wide-area surveys are less feasible. SALTUS is capable of mapping an area of $\approx$5\,arcmin$^2$ without repointing by using a fine-steering mirror. This field-of-view is very similar to the map sizes produced by single JWST/NIRCam or MIRI imaging pointings. In terms of resolution, sensitivity, and survey capabilities, SALTUS can be thought of as the far-IR analog of JWST, building on that mission's early success. 

The spectral mapping survey speed of SALTUS is comparable to that achievable with smaller 2,m-class apertures, largely because of the substantially better sensitivity of SALTUS. In comparison to surveys with SPICA/SAFARI, for example, SALTUS/SAFARI-Lite could survey a 1\,arcmin$^2$ area to the same depth in an equal amount of observing time, but with 30$\times$ more spatial resolution elements \citep{roelfsema18}. These capabilities lend themselves particularly well to making the first deep far-IR spectral surveys of deep extragalactic legacy fields (GOODS-N, GOODS-S, etc.). Even though SALTUS is not designed for survey-mode operations, its capabilities lend themselves well to JWST-like survey areas and speeds.

\section{Key High-Redshift Science Goals for SALTUS} \label{sci}

The high-redshift extragalactic science goals of SALTUS are designed to respond the the Decadal Survey’s “Cosmic Ecosystems” theme and address the key science questions identified for a far-IR probe-class mission. The high-redshift science goals are also highly complementary to observations planned for the nearby universe (see Levy \etal, subm.), which will provide a highly-resolved anchor for the more distant measurements we described. As a versatile, extremely sensitive observatory with arcsecond-level spatial resolution, SALTUS would be complementary to the near- and mid-IR capabilities of JWST. In this section we highlight an ambitious science campaign designed to answer outstanding questions raised by the Decadal Survey, including: How do gas, metals, and dust flow into, through, and out of galaxies? How do supermassive black holes form and how is their growth coupled to the growth of their host galaxies?

Thanks to the broad simultaneous wavelength coverage of the SAFARI-Lite instrument in particular, extragalactic observations of individual targets, pointed surveys of well-defined samples, and blank-field spectral mapping campaigns are all capable of addressing different aspects of the Cosmic Ecosystems science goals. Many of these goals (and beyond) will also be addressed by the wide variety of community GO observing programs enabled by the versatile and sensitive SALTUS capabilities.

\begin{figure}
\begin{center}
\includegraphics[width=0.75\textwidth]{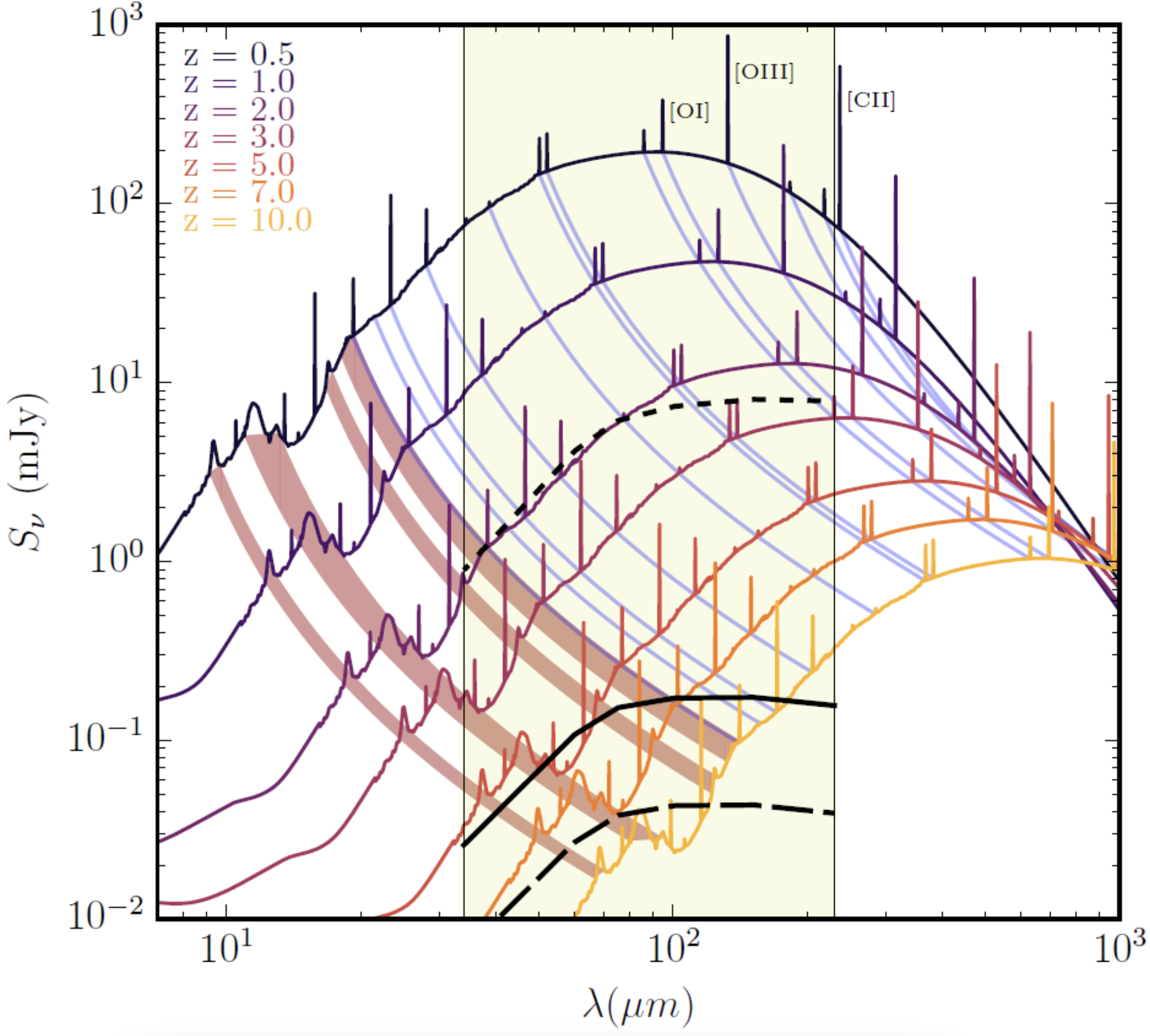}
\end{center}
\caption{ 
The spectral range of SALTUS over cosmic time. Schematic representation of the spectral energy distribution of a $3 \times 10^{12}$\,\Lsol star forming galaxy with redshift. Lines important to the science case and PAH features are traced through redshift, and dominant cooling lines (\oi, \oiii, \cii) are labeled. Out to $z\sim3$, SAFARI-Lite probes the peak of the dust continuum and the bulk of the dust emission. Beyond $z\sim3$, SAFARI-Lite takes over from JWST/MIRI to probe the red-shifted mid-IR PAH emission features. The yellow color-coded region indicates the wavelength range of SAFARI-Lite. The lower solid black curve is the detection limit for SAFARI-Lite at $R=300$ for pointed observations (1 hour, 5$\sigma$). The lower long-dashed line approximates a detection limit for wide PAH features, which span many channels. The short-dashed line is the SAFARI-Lite detection limit in mapping mode (1\,arcmin$^2$ area mapped in 1-hr at 5$\sigma$).
}\label{fig:specdpm}
\end{figure}

\subsection{Measure the rise of small dust grains in the early universe} \label{pahs}

The rapid expansion of our capabilities in the infrared over the last two decades has led to the key discovery that the bulk of active processes in the Universe -- evolving interstellar medium (ISM) chemistry and physics, star formation, black hole accretion -- is moderately to heavily obscured by interstellar dust even to high redshifts \citep[e.g.][]{madau14,zavala21}. This association is more than cosmetic: small-grain cosmic dust plays a key role in the photoelectric heating of gas in the ISM \citep[e.g.][]{tielens85} and the formation of molecular hydrogen \citep{bauschlicher98,foley18}, both of which may give rise to a fundamental relationship between dust and the birth of new stars \citep[e.g.][]{regan06,pope13,cortzen19}. 

The rise of cosmic dust in the early Universe is therefore a key area of focus in the study of galactic ecosystems. The wealth of diagnostics in a mid-IR galaxy spectrum includes several features arising from small, carbon-rich dust grains termed polycyclic aromatic hydrocarbons (PAHs), essentially the only spectral features of dust grains accessible at large distances. The strengths and ratios of these features are highly sensitive to their surrounding conditions (i.e. metallicity, radiation field, gas temperature). 

JWST is already providing tantalizing hints at early small grain dust enrichment: recently, the earliest direct detection of a PAH spectral feature at 3.3\,\um was made at $z=4.2$ ($\sim$1.5 Gyr after the Big Bang; \cite{spilker23}) and a UV feature (the so-called 2175\AA\ bump) that arises from carbon-rich grains has been observed at $z\gtrsim7$ ($\sim$800 Myr after the Big Bang; \cite{witstok23,markov24}).  These observations, combined with the high redshift (large grain) dust-rich galaxies revealed with ALMA, challenge models of dust production and survival which limit significant dust build-up to late times ($>$1.5 Gyr post Big Bang) via AGB star populations \citep[e.g.][]{michalowski15}.  The nature of alternative rapid dust producers will be quantified by JWST at lower redshifts, which has already mapped carbon-rich dust shells around Wolf-Rayet stars \citep{lau22} and in supernova ejecta \citep{shahbandeh24} at extremely high spatial resolution.

\begin{figure}
\begin{center}
\includegraphics[width=0.45\textwidth]{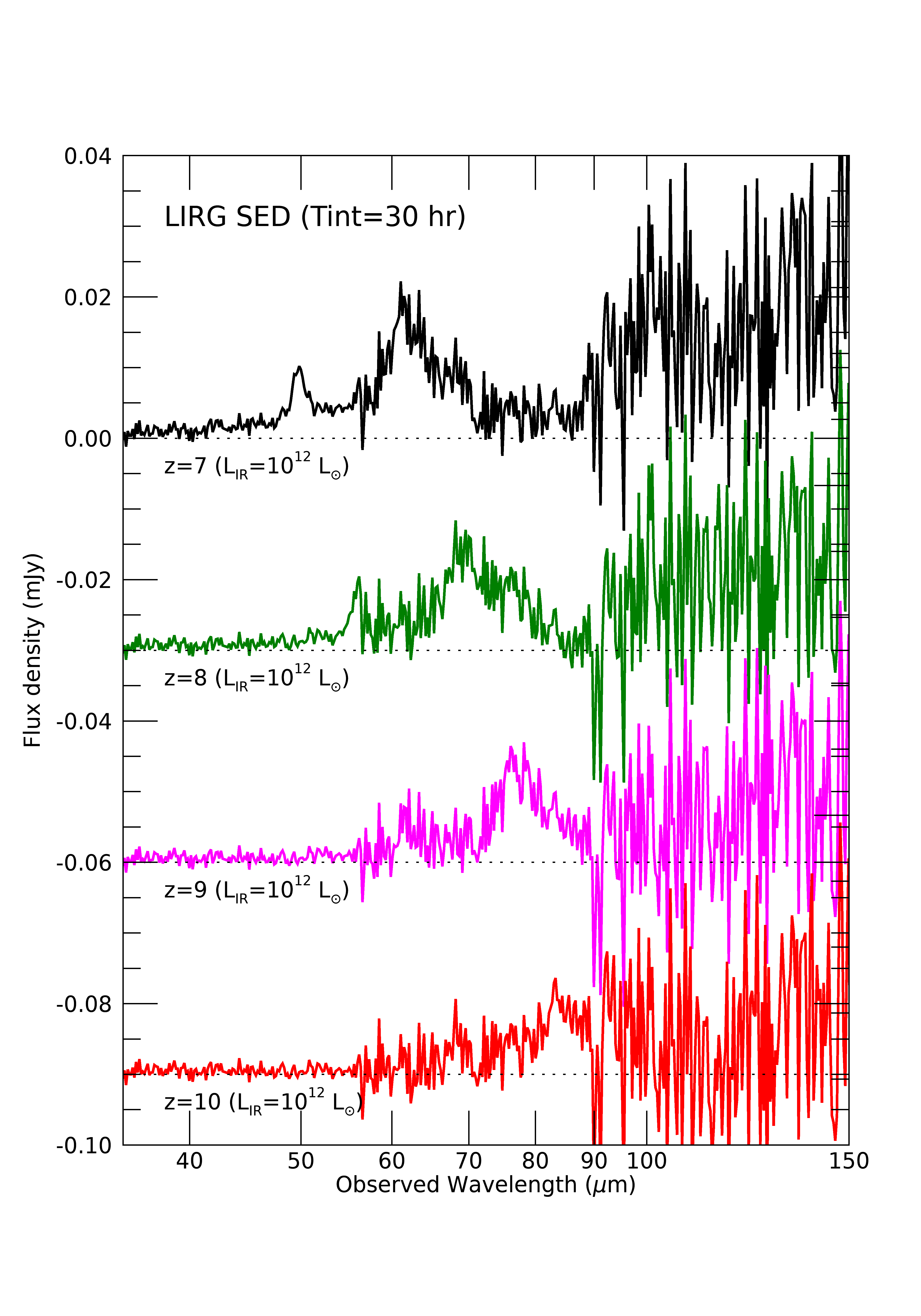}
\includegraphics[width=0.45\textwidth]{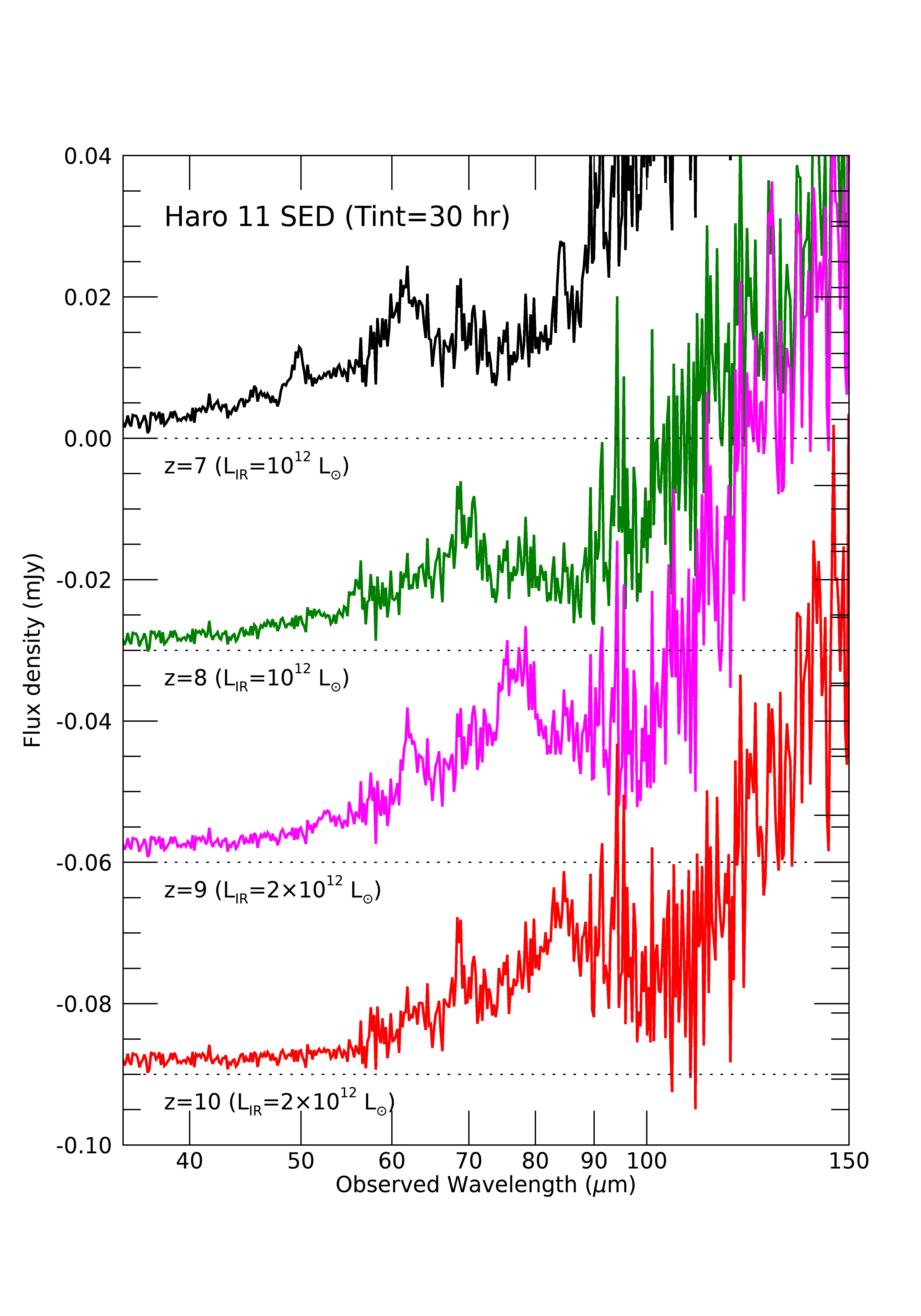}
\end{center}
\caption{ 
Simulated SAFARI-Lite PAH spectra for $z \ge 7$ galaxies, for both a dust- and metal-rich LIRG template (left) and a lower-metallicity template (Haro~11, 25\% solar; right). We scale each template to the total \lir listed by each curve, and simulate a 30\,hour integration. SALTUS can plausibly detect PAH emission even at $z>10$. Small carbon grains have already been detected in the UV in this redshift range by JWST \citep{witstok23,markov24}, and SALTUS will reveal their origin and properties by measuring PAH band ratios. At lower redshifts all PAH features are easily detected.
}\label{fig:pahseichii}
\end{figure}

JWST’s (and ALMA’s) role in directly developing the paradigm of early cosmic dust is limited, however, as neither has the wavelength range to access the rich diagnostic power of the mid-infrared (JWST can only observe one PAH feature at $z\gtrsim3$). SAFARI-Lite on SALTUS will have the wavelength coverage, spectral resolution, and sensitivity to detect the multiple PAH features in star forming galaxies at $z\sim7$ and beyond (Fig.~\ref{fig:pahseichii}). 

PAHs play a large role in balancing the heating and cooling processes in the ISM and environments for star formation by catalyzing the formation of H$_2$ molecules. To extract this information, we need to measure the strengths, equivalent widths, and feature ratios, all of which contain clues to the composition, grain sizes, and ionization states of the molecules \citep[e.g.][]{li01,draine21,maragkoudakis20,egorov23}. Up to $z\approx5.6$, SAFARI-Lite will simultaneously observe PAH features and the [Si{\scriptsize II}] 34.8\,\um emission line \citep[e.g.][]{mckinney21}. This line is a significant cooling channel in the strong radiation fields expected at high redshift \citep{kaufman06}, tracing the cycle of gas heating and cooling in photo-dissociation regions. 

As shown in Fig.~\ref{fig:pahseichii}, SALTUS/SAFARI-Lite is capable of measuring the integrated luminosity of the 6.2, 7.7, 8.6, 11.3, and 12.7\,\um PAH features at useful significance even at $z=10$ (cosmic age $<$500\,Myr). The expected rapid evolution of interstellar dust in galaxies through the first billion years will be charted through such spectra. In particular we note that the brightest of the PAH features do not redshift into the noisier long-wavelength channel of SAFARI-Lite until beyond $z>10$. At lower redshifts $3.5 < z < 7$, SALTUS' sensitivity allows even the faint 17\,\um PAH feature to be detected with ease.

\subsection{Measure the co-evolution of galaxy and black hole growth} \label{sfrbhar}

The two dominant processes occurring in active, evolving galaxies -- the creation of new stars and the accretion of material onto supermassive black holes -- are now understood to progress in parallel, with stellar mass and black hole growth rising to a peak during the epoch of cosmic noon before declining to present day \citep{madau14}. This parallel growth may be fundamental: feedback from black holes ubiquitous in galactic nuclei is widely invoked to regulate the evolution of massive galaxies, a scenario supported by both local observation and theoretical models \citep[e.g.][]{hopkins08,ceverino15,veilleux20}. The ubiquity of this pathway, however, is uncertain locally and both its efficacy and prevalence are even less constrained at higher redshifts \citep[e.g.][]{pacucci23,stone24}.

Understanding the causal link between star formation and active galactic nuclei (AGN) over cosmic time is a primary goal of the `Cosmic Ecosystems' SALTUS science theme. The vastly different spatial scales of these processes (kiloparsec versus parsec) have always presented a challenge. The predominant barrier after decades of X-ray and optical studies is that the key phase in any co-evolution is heavily obscured by dust and thus partially or completely invisible to short wavelength observations. Breaking through this barrier requires high resolution, sensitive far-IR spectroscopy to obtain a direct, unbiased view of the role of AGN in shaping galaxy assembly.

\begin{figure}
\begin{center}
\includegraphics[width=0.75\textwidth]{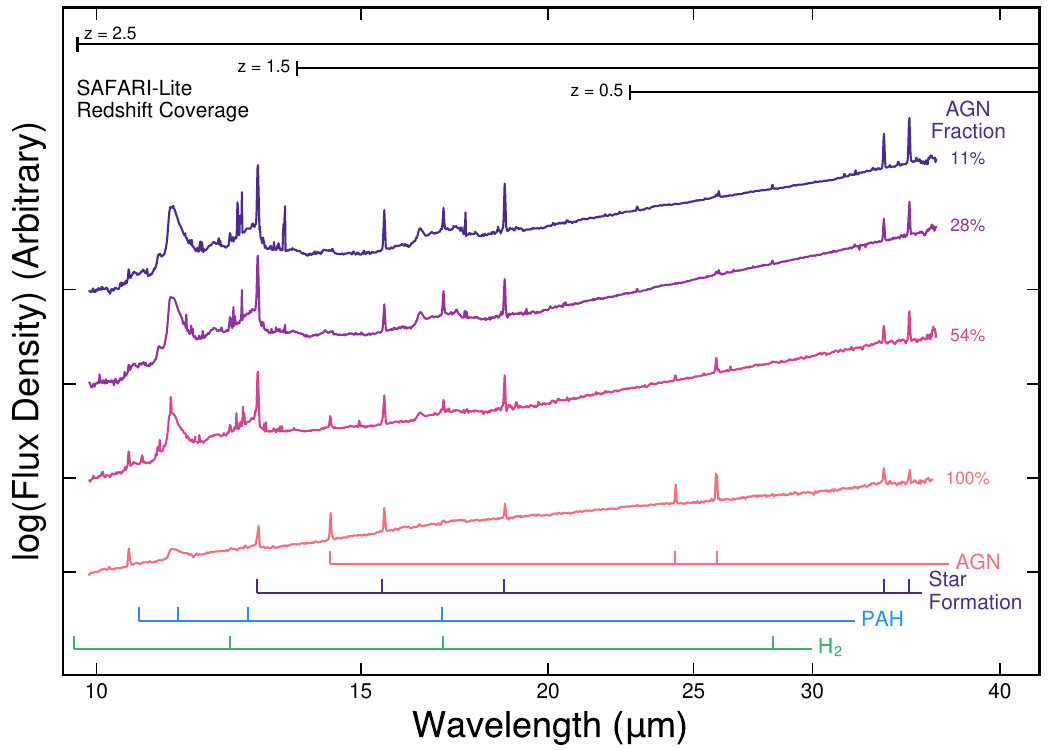}
\end{center}
\caption{ 
The rest-frame mid-IR is rich with diagnostics of star formation, AGN, galactic feedback/outflows, and shocks/warm H$_2$. For $z>0.5$ and extending beyond the peak of cosmic star formation, SAFARI-Lite will detect or constrain features probing each of these components in thousands of individual galaxies. Colored curves show template spectra for stacks of galaxies observed by Spitzer, ordered by the strength of the AGN contribution to the total luminosity \citep{stone22}.
}\label{fig:specjs}
\end{figure}

SAFARI-Lite will observe key diagnostics tracing star formation and black hole activity simultaneously in statistical samples of the galaxies that dominate stellar mass growth around cosmic noon ($z\sim1-4$) and in more luminous galaxies in the early Universe ($z>5$) (Figs.~\ref{fig:specdpm} and \ref{fig:specjs}). This is enabled by key far-IR emission lines that have been out of reach for all previous far-IR missions due to insufficient sensitivity or limited wavelength coverage, complementing JWST and ALMA at shorter and longer wavelengths. The fine structure lines \oiii 88\,\um and \cii 158\,\um will provide star formation rates to $z\sim1.6$, while the rest-frame mid-IR [Ne{\scriptsize II}] 12.8\,\um and [Ne{\scriptsize III}] 15.5\,\um are accessible from $z=1.6$ to beyond $z>10$. The high-excitation [Ne{\scriptsize V}]14 and 24\,\um, [Ne{\scriptsize VI}]7.6\,\um, and [O{\scriptsize IV}]26\,\um lines serve similar roles for black hole accretion rates \citep[e.g.][]{sajina22,stone22}.

This analysis made possible by SALTUS is the only pathway to link star formation and AGN during the dust obscured phase and establish the timing and thus relative importance of AGN activity in driving and/or quenching star formation. Simultaneously, SAFARI-Lite observations will provide a highly detailed analysis of the spectral features that reveal the suspected mechanism of this co-evolution: outflows driven by AGN feedback (Sec.~\ref{outflows}). This comprehensive approach will provide the breakthrough that drives our empirical and theoretical galaxy evolution models. 

\subsection{Measure galactic outflows across cosmic time}\label{outflows}

One of the most important realizations of the past two decades is the vital role that self-regulating feedback processes play in galaxy evolution. Feedback plays a crucial role in explaining many of the fundamental galaxy scaling relationships, including quenching of star formation in ``red and dead'' galaxies, the enrichment of heavy elements to great distances beyond a galaxy's stellar disk, and the connection between dark matter halos, star formation, stars, and supermassive black holes. Due to the large dynamic range of spatial and temporal scales over which it manifests, feedback remains the key unknown for cosmological simulations, resulting in a panoply of pre- and post-dictions for galaxy properties that differ by orders of magnitude.

Tracking the impact of feedback from an observational standpoint is a field in its nascency. Powerful galactic outflows of gas are clear feedback signatures and ubiquitously observed in the local universe, but detecting these outflows across cosmic time remains difficult even in the JWST era. In large part this is because the gas in outflows spans at least six orders of magnitude in both temperature and density \citep[e.g.][]{schneider17}, so it is difficult for any single observatory to detect and characterize a large portion of the outflowing material.

The far-IR wavelength regime uniquely offers access to outflow tracers ranging from the cold molecular phase (transitions of H$_2$O, OH), cool atomic gas (\cii 158\,\um, [Si{\scriptsize II}] 34.8\,\um, [Fe{\scriptsize II}] 26\,\um), and warm ionized gas ([S{\scriptsize III}] 33\,\um, \oiii 52 \& 88\,\um) that fall within the SAFARI-Lite bandpass from the nearby universe to Cosmic Noon, $z\sim2$ (Fig.~\ref{fig:specdpm}). In these spectral features, outflows manifest as high-velocity line wings (in emission, or, for the molecular phase, in absorption against the dust continuum). While the fastest outflows driven by supernova feedback will also be detectable, SALTUS’ science goal in this area is more narrowly focused on AGN-driven winds, which typically reach fast $>$1000\,\kms velocities distinguishable at the $R\sim300$ spectral resolution of SAFARI-Lite.\footnote{See also similar calculations performed for SPICA/SAFARI at the same spectral resolution; \cite{gonzalezalfonso17spica}.} 

\begin{figure}
\begin{center}
\includegraphics[width=0.9\textwidth]{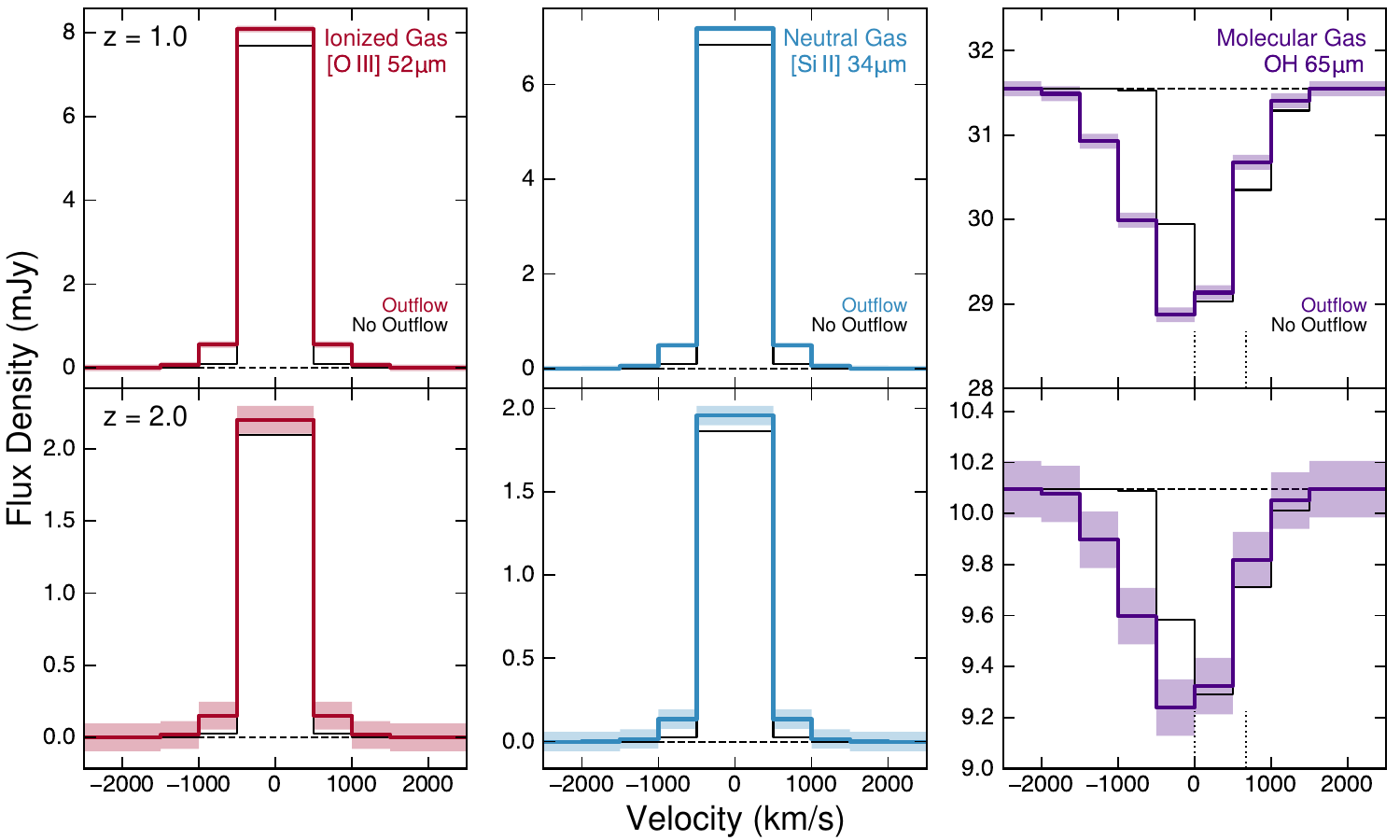}
\end{center}
\caption{ 
SALTUS/SAFARI-Lite will inventory galactic outflows with simultaneous observations of the cold molecular, cool atomic, and warm ionized gas phases detected using various far-IR spectral features. We show simulated spectra for a 1-hour observation of a $\lir = 2\times10^{12}$\,\Lsol galaxy with (thick colored lines) or without (thin black lines) a $\vout=700$\,\kms outflow. Colored shaded bands show the 1$\sigma$ uncertainty per channel. The broad outflow components are detected at both $z=1$ (top row) and $z=2$ (bottom row), especially in comparison to the no-outflow spectra.
}\label{fig:outflows}
\end{figure}

We illustrate the ability to recover multi-phase outflows using SALTUS/SAFARI-Lite observations in Figure~\ref{fig:outflows}. We created mock spectra at the (Nyquist-sampled) $R=300$ spectral resolution for a fiducial galaxy with $\lir = 2 \times 10^{12}$\,\Lsol at redshifts $z = 1.0,\,\,2.0$, either including or excluding an outflow component to the spectra. For the molecular phase, we simulated the OH 65\,\um doublet; for the neutral atomic, [Si{\scriptsize II}] 34.8\,\um; and for the ionized phase, \oiii 52\,\um. We used the IR spectral line scaling relations of \cite{spinoglio12} to define the non-outflow line fluxes. We assumed a nominal galaxy intrinsic line width of 400\,\kms, and a typical outflow velocity $\vout = 700$\,\kms. We assumed an OH absorption depth of $\approx$5\% of the continuum, similar to low- and high-redshift studies \citep[e.g.][]{gonzalezalfonso17,spilker20}, and for the emission lines we assumed a broad-wing amplitude 10\% of the line peak. Finally, we assumed a 1\,hr observation to determine the significance of the outflow component recovery.

Figure~\ref{fig:outflows} demonstrates that SALTUS will easily recover multi-phase outflow components in IR-luminous galaxies to Cosmic Noon and beyond. Even in a one-hour observation, broad line-wings and blueshifted absorption features are detected at very high significance at $z=1$ (implying that outflows could also be detected in lower-luminosity galaxies), and also recoverable at lower significance at $z=2$. We note especially that the outflow spectra are distinct from the same spectra that lack an outflow component, i.e. SALTUS can discern the presence vs. absence of outflows. SALTUS enables a comprehensive, multi-phase characterization of AGN-driven outflows that cannot be pieced together using existing or near-future observatories.

This science goal is highly complementary to the objectives outlined in Section~\ref{sfrbhar}. Multi-phase outflow tracers will be covered alongside key diagnostics of the star formation and black hole accretion rates commensally every time the telescope observes an extragalactic target, simultaneously detecting dust-immune metallicity indicators \citep[e.g.][]{rigopoulou18}. Over the mission lifetime, SALTUS will establish a complete library of galaxy star formation and black hole accretion rates with accompanying measurements of the outflow properties in the hot, cool, and cold gas phases in thousands of individual galaxies.

\subsection{Example Community Science in the JWST$+$SALTUS Era: Obscured AGN at Reionization}

One early lesson from JWST's first year is that rapid advances are made possible when large windows in discovery space are opened: unexpected galaxy populations are found for the first time, objects that were once thought to be rare are revealed to be common, and flawed assumptions from previous galaxy models are laid bare. Due to its sensitivity and lack of spatial/spectral confusion, SALTUS will allow a similar new discovery space to be opened, allowing the observatory to be nimble enough to address not just the science questions of the year 2024, but also the unknown questions that will be raised in the 2030s. We illustrate this responsiveness in the face of a recent unexpected JWST finding: that obscured AGN at very high redshifts may be far more common than previously expected.

Within the past year, JWST and ALMA observations have revealed an astonishing number of obscured or heavily-reddened AGN at $5<z<10$ \citep[e.g.][]{endsley23,furtak23,goulding23,greene23,kokorev23}. These spectroscopically-confirmed AGN are at least $\sim$30--100$\times$ more common than previously expected \citep[e.g.][]{shen20}, and therefore imply dramatic revisions to our understanding of early supermassive black hole growth. SALTUS is the only observatory capable of delivering much-needed constraints on the bolometric luminosities (and hence black hole growth rates) of these new, reionization-era AGN, as well as the origin of their dust obscuration. 

\begin{figure}
\begin{center}
\includegraphics[width=0.75\textwidth]{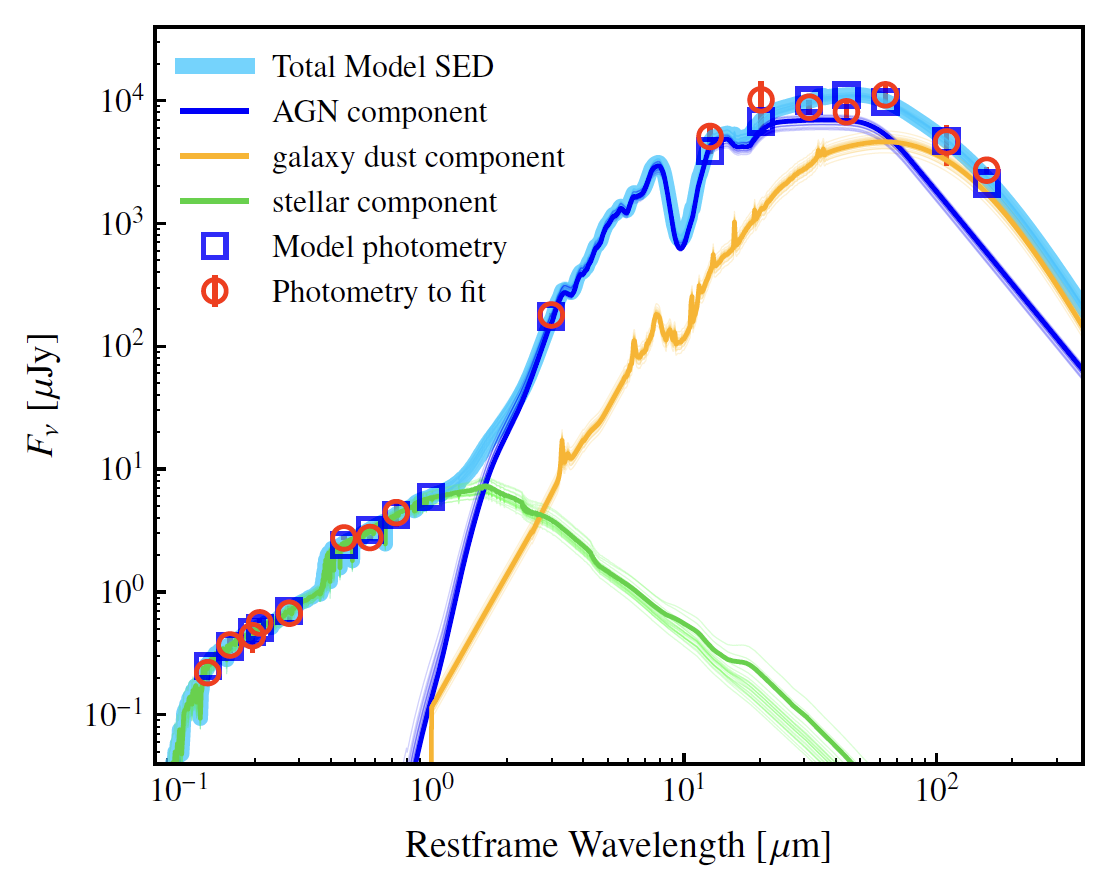}
\end{center}
\caption{ 
In its first year of operations, JWST has found an unexpected population of abundant obscured AGN at very high redshifts, including COSMOS-87259 shown here. SALTUS offers the ability to synergize with JWST discoveries into the 2030s thanks to its large aperture and lack of confusion limit. In this particular example, SALTUS would detect the mid-IR continuum and 9.7\,\um silicate feature (if present) in $<$1\,hour on-source. Reproduced with permission from \cite{endsley23}.
}\label{fig:agnendsley}
\end{figure}

We expect a single 10hour, 10arcmin$^2$ pointing with SALTUS to detect the mid-to-far infrared continua of approximately 10--30 obscured or heavily-reddened AGN at $5<z<10$ given recent number density measurements from JWST \citep{greene23}. These continuum measurements from SALTUS will provide our only direct probe of the warm and hot dust emission from the AGN torus around these very high-redshift AGN. Even the longest wavelength band with JWST/MIRI ($\sim$24\,\um) only probes up to rest-frame 3\,\um at $z=7$, far bluer than the peak of AGN dust emission (rest $\sim10-30$\,\um). Given the large diversity in mid-to-far infrared SED shapes of AGN \citep[e.g.][]{lyu17}, accurate constraints on the bolometric AGN luminosities, and hence black hole accretion rates, of these systems will only be possible by combining JWST data with measurements from SALTUS. Such data are critical to fill the huge hole in our census of very early black hole accretion density from this new and surprisingly abundant population of $z>5$ AGN. Targeted follow-up with SALTUS will yield a $\sim$10$\sigma$ spectroscopic continuum detection within minutes for the brightest known obscured AGN at $z>5$. The archetype example is COS-87259 which is confirmed to lie at $z=6.853$ within the 1.5\,deg$^2$ COSMOS field \citep{endsley23}. Several broadband photometric detections with Spitzer/MIPS, Herschel/PACS, and Herschel/SPIRE indicate that this obscured AGN has a flux density of approximately 0.3 to 10\,mJy that rises between observed-frame 34 to 230\,\um (Figure~\ref{fig:agnendsley}). The unprecedented mid- and far-infrared sensitivity and spectroscopic capabilities of SALTUS will not only dramatically improve constraints on the bolometric AGN luminosity of this system, but also enable the much-needed measurement of the rest-frame 9.7\,\um silicate absorption feature in this system. This silicate absorption feature is a highly valuable tracer of the amount of obscuration towards heavily-buried AGN like COS-87259, which cannot be detected in deep X-ray imaging \citep[e.g.][]{hickox18}. This feature also provides an estimate of the amount of obscuration due to the host galaxy ISM rather than circumnuclear dust. Such information is necessary to begin testing recent theoretical models that imply that the higher gas fractions and denser ISM of high-redshift galaxies will more readily lead to heavy AGN obscuration \citep[e.g.][]{trebitsch18,ni20}. COS-87259 was found within a relatively small field and we expect future surveys to rapidly identify many more similar systems \citep[e.g.][]{lyu23}. These will be excellent targets for pointed follow-up with SALTUS to deliver statistical constraints on the black hole growth rates, as well as extent and origin of obscuration among the most luminous, buried AGN in the reionization era.

We emphasize that this is only a single example drawn from the first year of JWST operations. Due to the confusion limits that have plagued small apertures in the far-IR, discovery space is limited only to those objects whose positions and redshifts were already known. In addition, detailed measurements to constrain the natures of the sources such as bolometric luminosities and relative role of star formation are severely compromised by confusion noise. SALTUS mitigates these limitations, enabling the discovery of previously-unknown galaxy populations and synergy with other long wavelength facilities including JWST.

\section{Conclusions} \label{conclusions}

From Section~\ref{perf}, it is clear that any future far-IR facility will offer transformative gains over previous observatories at this wavelength. Herschel's defining legacy for the distant universe is its large-area sensitive photometric mapping, but its relatively small aperture resulted in challenging problems from confusion noise that continue to be explored today. Herschel generally lacked the sensitivity for high-redshift spectroscopy, even in large samples of IR-luminous galaxies \citep{wardlow17,wilson17}. SALTUS would offer the first chance to probe the wealth of mid- and far-IR spectral diagnostics in galaxies throughout the universe, with no spatial or spectral confusion. Priority high-redshift measurements would detect small dust grains into the reionization epoch to measure the buildup of the dusty universe, simultaneously detect dust-immune indicators of star formation and black hole accretion in galaxies to Cosmic Noon and beyond, and measure the multiphase gas in fast galactic outflows that slow future galaxy growth and feed the circumgalactic medium. Its versatile capabilities ensure that SALTUS will also be responsive to the needs of the high-redshift community of the 2030s, after a decade of JWST operations, answering questions that JWST has only just begun to pose.

\subsection*{Disclosures}
The authors report no conflicts of interest.

\subsection*{Code and Data Availability}
No code or data were used in the preparation of this manuscript.



\begin{thebibliography}{10}

\bibitem{roelfsema18}
P.~R. {Roelfsema}, H.~{Shibai}, L.~{Armus}, {\em et~al.}, ``{SPICA-A Large
  Cryogenic Infrared Space Telescope: Unveiling the Obscured Universe},'' {\em
  \pasa} {\bf 35}, e030  (2018).

\bibitem{meixner19}
M.~{Meixner}, A.~{Cooray}, D.~{Leisawitz}, {\em et~al.}, ``{Origins Space
  Telescope Mission Concept Study Report},'' {\em arXiv e-prints} ,
  arXiv:1912.06213  (2019).

\bibitem{chiar06}
J.~E. {Chiar} and A.~G.~G.~M. {Tielens}, ``{Pixie Dust: The Silicate Features
  in the Diffuse Interstellar Medium},'' {\em \apj} {\bf 637}, 774--785
  (2006).

\bibitem{wang19}
S.~{Wang} and X.~{Chen}, ``{The Optical to Mid-infrared Extinction Law Based on
  the APOGEE, Gaia DR2, Pan-STARRS1, SDSS, APASS, 2MASS, and WISE Surveys},''
  {\em \apj} {\bf 877}, 116  (2019).

\bibitem{pereirasantaella17}
M.~{Pereira-Santaella}, D.~{Rigopoulou}, D.~{Farrah}, {\em et~al.},
  ``{Far-infrared metallicity diagnostics: application to local ultraluminous
  infrared galaxies},'' {\em \mnras} {\bf 470}, 1218--1232  (2017).

\bibitem{peng21}
B.~{Peng}, C.~{Lamarche}, G.~J. {Stacey}, {\em et~al.}, ``{Far-Infrared Line
  Diagnostics: Improving N/O Abundance Estimates for Dusty Galaxies},'' {\em
  \apj} {\bf 908}, 166  (2021).

\bibitem{spinoglio22}
L.~{Spinoglio}, J.~A. {Fern{\'a}ndez-Ontiveros}, M.~A. {Malkan}, {\em et~al.},
  ``{SOFIA Observations of Far-IR Fine-structure Lines in Galaxies to Measure
  Metallicity},'' {\em \apj} {\bf 926}, 55  (2022).

\bibitem{chen23}
Y.~{Chen}, T.~{Jones}, R.~{Sanders}, {\em et~al.}, ``{Accurate oxygen abundance
  of interstellar gas in Mrk 71 from optical and infrared spectra},'' {\em
  Nature Astronomy} {\bf 7}, 771--778  (2023).

\bibitem{chartab22}
N.~{Chartab}, A.~{Cooray}, J.~{Ma}, {\em et~al.}, ``{Low gas-phase
  metallicities of ultraluminous infrared galaxies are a result of dust
  obscuration},'' {\em Nature Astronomy} {\bf 6}, 844--849  (2022).

\bibitem{gonzalezalfonso14}
E.~{Gonz{\'a}lez-Alfonso}, J.~{Fischer}, J.~{Graci{\'a}-Carpio}, {\em et~al.},
  ``{The Mrk 231 molecular outflow as seen in OH},'' {\em \aap} {\bf 561}, A27
  (2014).

\bibitem{spilker20}
J.~S. {Spilker}, K.~A. {Phadke}, M.~{Aravena}, {\em et~al.}, ``{Ubiquitous
  Molecular Outflows in z > 4 Massive, Dusty Galaxies. I. Sample Overview and
  Clumpy Structure in Molecular Outflows on 500 pc Scales},'' {\em \apj} {\bf
  905}, 85  (2020).

\bibitem{herreracamus21}
R.~{Herrera-Camus}, N.~{F{\"o}rster Schreiber}, R.~{Genzel}, {\em et~al.},
  ``{Kiloparsec view of a typical star-forming galaxy when the Universe was
  {\ensuremath{\sim}}1 Gyr old. I. Properties of outflow, halo, and
  interstellar medium},'' {\em \aap} {\bf 649}, A31  (2021).

\bibitem{smercina18}
A.~{Smercina}, J.~D.~T. {Smith}, D.~A. {Dale}, {\em et~al.}, ``{After the Fall:
  The Dust and Gas in E+A Post-starburst Galaxies},'' {\em \apj} {\bf 855}, 51
  (2018).

\bibitem{draine21}
B.~T. {Draine}, A.~{Li}, B.~S. {Hensley}, {\em et~al.}, ``{Excitation of
  Polycyclic Aromatic Hydrocarbon Emission: Dependence on Size Distribution,
  Ionization, and Starlight Spectrum and Intensity},'' {\em \apj} {\bf 917}, 3
  (2021).

\bibitem{dole04b}
H.~{Dole}, G.~H. {Rieke}, G.~{Lagache}, {\em et~al.}, ``{Confusion of
  Extragalactic Sources in the Mid- and Far-Infrared: Spitzer and Beyond},''
  {\em \apjs} {\bf 154}, 93--96  (2004).

\bibitem{roseboom10}
I.~G. {Roseboom}, S.~J. {Oliver}, M.~{Kunz}, {\em et~al.}, ``{The Herschel
  Multi-Tiered Extragalactic Survey: source extraction and
  cross-identifications in confusion-dominated SPIRE images},'' {\em \mnras}
  {\bf 409}, 48--65  (2010).

\bibitem{hurley17}
P.~D. {Hurley}, S.~{Oliver}, M.~{Betancourt}, {\em et~al.}, ``{HELP: XID+, the
  probabilistic de-blender for Herschel SPIRE maps},'' {\em \mnras} {\bf 464},
  885--896  (2017).

\bibitem{liu18}
D.~{Liu}, E.~{Daddi}, M.~{Dickinson}, {\em et~al.},
  ``{{\textquotedblleft}Super-deblended{\textquotedblright} Dust Emission in
  Galaxies. I. The GOODS-North Catalog and the Cosmic Star Formation Rate
  Density out to Redshift 6},'' {\em \apj} {\bf 853}, 172  (2018).

\bibitem{madau14}
P.~{Madau} and M.~{Dickinson}, ``{Cosmic Star-Formation History},'' {\em \araa}
  {\bf 52}, 415--486  (2014).

\bibitem{zavala21}
J.~A. {Zavala}, C.~M. {Casey}, S.~M. {Manning}, {\em et~al.}, ``{The Evolution
  of the IR Luminosity Function and Dust-obscured Star Formation over the Past
  13 Billion Years},'' {\em \apj} {\bf 909}, 165  (2021).

\bibitem{tielens85}
A.~G.~G.~M. {Tielens} and D.~{Hollenbach}, ``{Photodissociation regions. I.
  Basic model.},'' {\em \apj} {\bf 291}, 722--746  (1985).

\bibitem{bauschlicher98}
J.~{Bauschlicher}, Charles~W., ``{The Reaction of Polycyclic Aromatic
  Hydrocarbon Cations with Hydrogen Atoms: The Astrophysical Implications},''
  {\em \apjl} {\bf 509}, L125--L127  (1998).

\bibitem{foley18}
N.~{Foley}, S.~{Cazaux}, D.~{Egorov}, {\em et~al.}, ``{Molecular hydrogen
  formation on interstellar PAHs through Eley-Rideal abstraction reactions},''
  {\em \mnras} {\bf 479}, 649--656  (2018).

\bibitem{regan06}
M.~W. {Regan}, M.~D. {Thornley}, S.~N. {Vogel}, {\em et~al.}, ``{The Radial
  Distribution of the Interstellar Medium in Disk Galaxies: Evidence for
  Secular Evolution},'' {\em \apj} {\bf 652}, 1112--1121  (2006).

\bibitem{pope13}
A.~{Pope}, J.~{Wagg}, D.~{Frayer}, {\em et~al.}, ``{Probing the Interstellar
  Medium of z \raisebox{-0.5ex}\textasciitilde 1 Ultraluminous Infrared
  Galaxies through Interferometric Observations of CO and Spitzer Mid-infrared
  Spectroscopy},'' {\em \apj} {\bf 772}, 92  (2013).

\bibitem{cortzen19}
I.~{Cortzen}, J.~{Garrett}, G.~{Magdis}, {\em et~al.}, ``{PAHs as tracers of
  the molecular gas in star-forming galaxies},'' {\em \mnras} {\bf 482},
  1618--1633  (2019).

\bibitem{spilker23}
J.~S. {Spilker}, K.~A. {Phadke}, M.~{Aravena}, {\em et~al.}, ``{Spatial
  variations in aromatic hydrocarbon emission in a dust-rich galaxy},'' {\em
  \nat} {\bf 618}, 708--711  (2023).

\bibitem{witstok23}
J.~{Witstok}, I.~{Shivaei}, R.~{Smit}, {\em et~al.}, ``{Carbonaceous dust
  grains seen in the first billion years of cosmic time},'' {\em \nat} {\bf
  621}, 267--270  (2023).

\bibitem{markov24}
V.~{Markov}, S.~{Gallerani}, A.~{Ferrara}, {\em et~al.}, ``{Dust attenuation
  evolution in $z \sim 2$-$12$ JWST galaxies},'' {\em arXiv e-prints} ,
  arXiv:2402.05996  (2024).

\bibitem{michalowski15}
M.~J. {Micha{\l}owski}, ``{Dust production 680-850 million years after the Big
  Bang},'' {\em \aap} {\bf 577}, A80  (2015).

\bibitem{lau22}
R.~M. {Lau}, M.~J. {Hankins}, Y.~{Han}, {\em et~al.}, ``{Nested dust shells
  around the Wolf-Rayet binary WR 140 observed with JWST},'' {\em Nature
  Astronomy} {\bf 6}, 1308--1316  (2022).

\bibitem{shahbandeh24}
M.~{Shahbandeh}, C.~{Ashall}, P.~{Hoeflich}, {\em et~al.}, ``{JWST NIRSpec+MIRI
  Observations of the nearby Type IIP supernova 2022acko},'' {\em arXiv
  e-prints} , arXiv:2401.14474  (2024).

\bibitem{li01}
A.~{Li} and B.~T. {Draine}, ``{Infrared Emission from Interstellar Dust. II.
  The Diffuse Interstellar Medium},'' {\em \apj} {\bf 554}, 778--802  (2001).

\bibitem{maragkoudakis20}
A.~{Maragkoudakis}, E.~{Peeters}, and A.~{Ricca}, ``{Probing the size and
  charge of polycyclic aromatic hydrocarbons},'' {\em \mnras} {\bf 494},
  642--664  (2020).

\bibitem{egorov23}
O.~V. {Egorov}, K.~{Kreckel}, K.~M. {Sandstrom}, {\em et~al.}, ``{PHANGS-JWST
  First Results: Destruction of the PAH Molecules in H II Regions Probed by
  JWST and MUSE},'' {\em \apjl} {\bf 944}, L16  (2023).

\bibitem{mckinney21}
J.~{McKinney}, L.~{Armus}, A.~{Pope}, {\em et~al.}, ``{Regulating Star
  Formation in Nearby Dusty Galaxies: Low Photoelectric Efficiencies in the
  Most Compact Systems},'' {\em \apj} {\bf 908}, 238  (2021).

\bibitem{kaufman06}
M.~J. {Kaufman}, M.~G. {Wolfire}, and D.~J. {Hollenbach}, ``{[Si II], [Fe II],
  [C II], and H$_{2}$ Emission from Massive Star-forming Regions},'' {\em \apj}
  {\bf 644}, 283--299  (2006).

\bibitem{hopkins08}
P.~F. {Hopkins}, L.~{Hernquist}, T.~J. {Cox}, {\em et~al.}, ``{A Cosmological
  Framework for the Co-Evolution of Quasars, Supermassive Black Holes, and
  Elliptical Galaxies. I. Galaxy Mergers and Quasar Activity},'' {\em \apjs}
  {\bf 175}, 356--389  (2008).

\bibitem{ceverino15}
D.~{Ceverino}, A.~{Dekel}, D.~{Tweed}, {\em et~al.}, ``{Early formation of
  massive, compact, spheroidal galaxies with classical profiles by violent disc
  instability or mergers},'' {\em \mnras} {\bf 447}, 3291--3310  (2015).

\bibitem{veilleux20}
S.~{Veilleux}, R.~{Maiolino}, A.~D. {Bolatto}, {\em et~al.}, ``{Cool outflows
  in galaxies and their implications},'' {\em \aapr} {\bf 28}, 2  (2020).

\bibitem{pacucci23}
F.~{Pacucci}, B.~{Nguyen}, S.~{Carniani}, {\em et~al.}, ``{JWST CEERS and JADES
  Active Galaxies at z = 4-7 Violate the Local M $_{{\textbullet}}$-M
  $_{{\ensuremath{\star}}}$ Relation at >3{\ensuremath{\sigma}}: Implications
  for Low-mass Black Holes and Seeding Models},'' {\em \apjl} {\bf 957}, L3
  (2023).

\bibitem{stone24}
M.~A. {Stone}, J.~{Lyu}, G.~H. {Rieke}, {\em et~al.}, ``{Undermassive Host
  Galaxies of Five z {\ensuremath{\sim}} 6 Luminous Quasars Detected with
  JWST},'' {\em \apj} {\bf 964}, 90  (2024).

\bibitem{sajina22}
A.~{Sajina}, M.~{Lacy}, and A.~{Pope}, ``{The Past and Future of Mid-Infrared
  Studies of AGN},'' {\em Universe} {\bf 8}, 356  (2022).

\bibitem{stone22}
M.~{Stone}, A.~{Pope}, J.~{McKinney}, {\em et~al.}, ``{Measuring Star Formation
  and Black Hole Accretion Rates in Tandem Using Mid-infrared Spectra of Local
  Infrared Luminous Galaxies},'' {\em \apj} {\bf 934}, 27  (2022).

\bibitem{schneider17}
E.~E. {Schneider} and B.~E. {Robertson}, ``{Hydrodynamical Coupling of Mass and
  Momentum in Multiphase Galactic Winds},'' {\em \apj} {\bf 834}, 144  (2017).

\bibitem{gonzalezalfonso17spica}
E.~{Gonz{\'a}lez-Alfonso}, L.~{Armus}, F.~J. {Carrera}, {\em et~al.},
  ``{Feedback and Feeding in the Context of Galaxy Evolution with SPICA: Direct
  Characterisation of Molecular Outflows and Inflows},'' {\em \pasa} {\bf 34},
  e054  (2017).

\bibitem{spinoglio12}
L.~{Spinoglio}, K.~M. {Dasyra}, A.~{Franceschini}, {\em et~al.},
  ``{Far-IR/Submillimeter Spectroscopic Cosmological Surveys: Predictions of
  Infrared Line Luminosity Functions for z < 4 Galaxies},'' {\em \apj} {\bf
  745}, 171  (2012).

\bibitem{gonzalezalfonso17}
E.~{Gonz{\'a}lez-Alfonso}, J.~{Fischer}, H.~W.~W. {Spoon}, {\em et~al.},
  ``{Molecular Outflows in Local ULIRGs: Energetics from Multitransition OH
  Analysis},'' {\em \apj} {\bf 836}, 11  (2017).

\bibitem{rigopoulou18}
D.~{Rigopoulou}, M.~{Pereira-Santaella}, G.~E. {Magdis}, {\em et~al.}, ``{On
  the far-infrared metallicity diagnostics: applications to high-redshift
  galaxies},'' {\em \mnras} {\bf 473}, 20--29  (2018).

\bibitem{endsley23}
R.~{Endsley}, D.~P. {Stark}, J.~{Lyu}, {\em et~al.}, ``{ALMA confirmation of an
  obscured hyperluminous radio-loud AGN at z = 6.853 associated with a dusty
  starburst in the 1.5 deg$^{2}$ COSMOS field},'' {\em \mnras} {\bf 520},
  4609--4620  (2023).

\bibitem{furtak23}
L.~J. {Furtak}, A.~{Zitrin}, A.~{Plat}, {\em et~al.}, ``{JWST UNCOVER:
  Extremely Red and Compact Object at zphot ~ 7.6 Triply Imaged by A2744},''
  {\em \apj} {\bf 952}, 142  (2023).

\bibitem{goulding23}
A.~D. {Goulding}, J.~E. {Greene}, D.~J. {Setton}, {\em et~al.}, ``{UNCOVER: The
  Growth of the First Massive Black Holes from JWST/NIRSpec-Spectroscopic
  Redshift Confirmation of an X-Ray Luminous AGN at z = 10.1},'' {\em \apjl}
  {\bf 955}, L24  (2023).

\bibitem{greene23}
J.~E. {Greene}, I.~{Labbe}, A.~D. {Goulding}, {\em et~al.}, ``{UNCOVER
  spectroscopy confirms a surprising ubiquity of AGN in red galaxies at
  $z>5$},'' {\em arXiv e-prints} , arXiv:2309.05714  (2023).

\bibitem{kokorev23}
V.~{Kokorev}, S.~{Fujimoto}, I.~{Labbe}, {\em et~al.}, ``{UNCOVER: A NIRSpec
  Identification of a Broad-line AGN at z = 8.50},'' {\em \apjl} {\bf 957}, L7
  (2023).

\bibitem{shen20}
X.~{Shen}, P.~F. {Hopkins}, C.-A. {Faucher-Gigu{\`e}re}, {\em et~al.}, ``{The
  bolometric quasar luminosity function at z = 0-7},'' {\em \mnras} {\bf 495},
  3252--3275  (2020).

\bibitem{lyu17}
J.~{Lyu}, G.~H. {Rieke}, and Y.~{Shi}, ``{Dust-deficient Palomar-Green Quasars
  and the Diversity of AGN Intrinsic IR Emission},'' {\em \apj} {\bf 835}, 257
  (2017).

\bibitem{hickox18}
R.~C. {Hickox} and D.~M. {Alexander}, ``{Obscured Active Galactic Nuclei},''
  {\em \araa} {\bf 56}, 625--671  (2018).

\bibitem{trebitsch18}
M.~{Trebitsch}, M.~{Volonteri}, Y.~{Dubois}, {\em et~al.}, ``{Escape of
  ionizing radiation from high-redshift dwarf galaxies: role of AGN
  feedback},'' {\em \mnras} {\bf 478}, 5607--5625  (2018).

\bibitem{ni20}
Y.~{Ni}, T.~{Di Matteo}, R.~{Gilli}, {\em et~al.}, ``{QSO obscuration at high
  redshift (z {\ensuremath{\gtrsim}} 7): predictions from the BLUETIDES
  simulation},'' {\em \mnras} {\bf 495}, 2135--2151  (2020).

\bibitem{lyu23}
J.~{Lyu}, S.~{Alberts}, G.~H. {Rieke}, {\em et~al.}, ``{AGN Selection and
  Demographics: A New Age with JWST/MIRI},'' {\em arXiv e-prints} ,
  arXiv:2310.12330  (2023).

\bibitem{wardlow17}
J.~L. {Wardlow}, A.~{Cooray}, W.~{Osage}, {\em et~al.}, ``{The Interstellar
  Medium in High-redshift Submillimeter Galaxies as Probed by Infrared
  Spectroscopy*},'' {\em \apj} {\bf 837}, 12  (2017).

\bibitem{wilson17}
D.~{Wilson}, A.~{Cooray}, H.~{Nayyeri}, {\em et~al.}, ``{Stacked Average
  Far-infrared Spectrum of Dusty Star-forming Galaxies from the Herschel/SPIRE
  Fourier Transform Spectrometer},'' {\em \apj} {\bf 848}, 30  (2017).

\end{thebibliography}

\vspace{2ex}\noindent\textbf{Justin Spilker} is an assistant professor in the Department of Physics and Astronomy and the Mitchell Institute of Texas A\&M University. 

\vspace{2ex}\noindent\textbf{Rebecca C. Levy} is a NSF Astronomy \& Astrophysics Postdoctoral Fellow at the University of Arizona.

\vspace{2ex}\noindent\textbf{Daniel P. Marrone} is a professor of astronomy at the University of Arizona.

\vspace{2ex}\noindent\textbf{Stacey Alberts} is an assistant professor at the University of Arizona.

\vspace{2ex}\noindent\textbf{Mark Dickinson} is an astronomer at NSF's National Optical-Infrared Astronomy Research Laboratory (NOIRLab). 

\vspace{2ex}\noindent\textbf{George Rieke} is a Regents' Professor of Astronomy and Planetary Sciences at the University of Arizona.

\vspace{2ex}\noindent\textbf{Alexander Tielens} is a professor of astronomy in the Astronomy Department of the University of Maryland, College Park.

\vspace{2ex}\noindent\textbf{Christopher K. Walker} is a professor of astronomy at the University of Arizona, and the principal investigator of the SALTUS mission concept.

\vspace{1ex}
\noindent Biographies of the other authors are not available.


\end{spacing}
\end{document}